\RequirePackage{fix-cm}
\documentclass[twocolumn,epjc3]{svjour3}  
\usepackage[numbers,square,comma,sort&compress]{natbib}
\usepackage[utf8]{inputenc}
\usepackage{graphicx}  
\usepackage{latexsym}   
\usepackage{enumerate}
\usepackage{rotating,booktabs,multirow}
\usepackage{amsmath}
\usepackage{amsfonts}
\usepackage{amssymb}
\usepackage{multirow}
\usepackage{bm}
\usepackage{xcolor}
\usepackage{hyperref}

\journalname{Eur. Phys. J. C}

\begin{document}
\sloppy

\title{Separating Equation-of-State Dynamics from Hadronic Rescattering in Low-Mass Dileptons
}

\author{Apiwit Kittiratpattana\thanksref{addr1}
        \and
        Ayut Limphirat\thanksref{addr1}
        \and
        Yupeng Yan\thanksref{addr1}
        \and
        Christoph Herold\thanksref{e1,addr1}
}

\thankstext{e1}{e-mail: herold@g.sut.ac.th}


\institute{Center of Excellence in High Energy Physics \& Astrophysics, School of Physics, Suranaree University of Technology, University Avenue 111, Nakhon Ratchasima 30000, Thailand \label{addr1}
}

\date{Received: date / Accepted: date}

\maketitle

\begin{abstract}
We investigate low-mass dilepton emission as a probe of the QCD equation of state and phase structure within non-equilibrium chiral fluid dynamics, comparing first-order phase-transition and crossover scenarios at 
$\sqrt{s_{\rm NN}}=2.20-6.20$~GeV. To disentangle effects of the macroscopic equation-of-state dynamics from conventional hadronic in-medium modifications, the $\rho$ and $\omega$ meson self-energies are calculated from the same resonance-driven forward-scattering amplitudes in both scenarios. We find two temporally distinct signatures of the first-order phase transition: an early enhancement in the vector-meson pole region associated with the non-equilibrium evolution through the phase transition, and a later enhancement of the low-mass continuum driven by reheating and the prolonged fireball evolution. Both effects survive integration over the complete space-time evolution, with the pole-mass region retaining the strongest sensitivity to the phase structure. Across the investigated beam energies, the pole-mass excitation function retains a pronounced sensitivity to first-order transition dynamics at low collision energies, identifying this mass region as a promising target for future dilepton beam-energy scans.
\end{abstract}

\section{Introduction}
\label{sec:intro}

One of the central goals of heavy-ion physics is to understand the equation of state (EoS) of strongly interacting matter and to determine whether a first-order phase transition (FOPT) and a critical endpoint exist in the QCD phase diagram.
At vanishing baryon chemical potential $\mu_B$, lattice QCD has established a smooth crossover (COV) \cite{Aoki:2006we,Aoki:2006br,Aoki:2009sc,Borsanyi:2010bp,Borsanyi:2010cj,HotQCD:2014kol}.
However, at large baryon densities, the fermion sign problem prevents direct lattice calculations, limiting first-principle constraints on the high-density EoS~\cite{deForcrand:2009zkb,Nagata:2021ugx}.
The beam-energy range covered by HADES and the upcoming CBM experiment at FAIR is specifically designed to probe this region, reaching the highest net-baryon densities achievable in the
laboratory~\cite{HADES:2020wpc,CBM:2016kpk}.
Measurements in this regime provide experimental constraints on the dense QCD EoS, which also determines the structure, stability, and mergers of neutron stars~\cite{Miller:2019cac,Riley:2019yda,Miller:2021qha,Riley:2021pdl}. Heavy-ion collisions therefore offer a complementary laboratory for studying the same dense matter probed through gravitational-wave and multi-messenger observations~\cite{LIGOScientific:2017vwq,Huth:2021bsp,Most:2022wgo,OmanaKuttan:2022the}.

Dileptons are particularly promising probes of the QCD phase structure \cite{Seck:2020qbx,Savchuk:2022aev}. Since they escape the strongly interacting medium without significant final-state interactions after production, their invariant mass spectra preserve information from different stages of the fireball evolution, allowing the time history of the medium to be investigated~\cite{McLerran:1984ay,Gale:1990pn,Rapp:2014hha}.
In the low invariant mass region, dilepton production is dominated by electromagnetic decays of vector mesons, making it directly sensitive to in-medium modifications of their spectral functions~\cite{Rapp:1999us,Brown:1991kk,vanHees:2006ng,vanHees:2007th}.
Two main theoretical approaches have been developed to describe these modifications.
In the many-body approach of Rapp and Wambach~\cite{Rapp:1999ej}, the $\rho$ spectral function is broadened through hadronic scattering at finite temperature and baryon density. The resulting rates have been applied on top of a coarse-grained UrQMD~\cite{Endres:2014zua,Endres:2015fna,Galatyuk:2015pkq} or SMASH~\cite{Staudenmaier:2017vtq} evolution in which the bulk dynamics is purely hadronic.
Alternatively, PHSD propagates off-shell spectral functions within a dynamical quasiparticle background matched to lattice QCD, incorporating a deconfinement transition and aspects of chiral symmetry restoration, while the low-mass $\rho$ broadening remains driven by collisional hadronic interactions~\cite{Linnyk:2015rco,Jorge:2025wwp}.
Both approaches successfully describe the observed low-mass dilepton enhancement.

Previous studies have demonstrated that dilepton emission can be sensitive to the presence of a FOPT and to the associated modification of the fireball lifetime and thermodynamic evolution~\cite{Seck:2020qbx,Savchuk:2022aev}. However, dilepton spectra in the low-mass region are simultaneously shaped by strong hadronic in-medium effects, in particular the broadening of the $\rho$ and $\omega$ spectral functions through interactions with the surrounding medium. This makes it important to disentangle modifications caused by the macroscopic EoS dynamics from those arising from the microscopic hadronic emission rates. The present work addresses this separation explicitly. We employ the same resonance-driven in-medium vector-meson self-energies in the FOPT and COV scenarios while varying the underlying chiral phase structure and its resulting dynamical evolution. This controlled comparison allows differences in the dilepton spectra to be traced directly to the EoS-driven space-time evolution. Moreover, by resolving the emission in time and invariant mass, we identify distinct dynamical origins of the resulting signatures: an early modification of the vector-meson pole region associated with the non-equilibrium evolution through the first-order transition, and a later enhancement of the low-mass continuum associated with reheating and the prolonged fireball evolution. We finally investigate how these temporally distinct effects survive in the integrated spectra and their beam-energy dependence and ultimately translate into experimentally accessible observables.

In this work, we combine non-equilibrium chiral fluid dynamics (N$\chi$FD)~\cite{Nahrgang:2011mg,Herold:2018ptm,Yusoff:2025cvu} with dilepton emission rates based on vector-meson self-energies~\cite{Eletsky:2001bb} to study dilepton production in the energy range $\sqrt{s_{\rm NN}}=2.20-6.20$~GeV. The chiral phase structure is varied through the pion mass to generate FOPT and COV scenarios, while the microscopic hadronic scattering channels entering the $\rho$ and $\omega$ meson self-energies are kept identical. The resonance content of the forward scattering amplitudes is extended according to the UrQMD/PDG particle table~\cite{Bass:1998ca,Bleicher:1999xi}. 

The remainder of this paper is organized as follows.
Section~\ref{sec:nxfd} introduces the N$\chi$FD framework.
Section~\ref{sec:selfenergy} presents the vector meson self-energies from
forward scattering amplitudes and the dilepton emission rate.
Sections~\ref{sec:timeprofile}--\ref{sec:excitation} present results for the
emission time profiles, invariant mass spectra, and excitation function. The results are summarized and discussed in Section~\ref{sec:discussion}.

\section{Non-equilibrium chiral fluid dynamics}
\label{sec:nxfd}

\begin{figure*}[t]
\centering
\includegraphics[width=\textwidth]{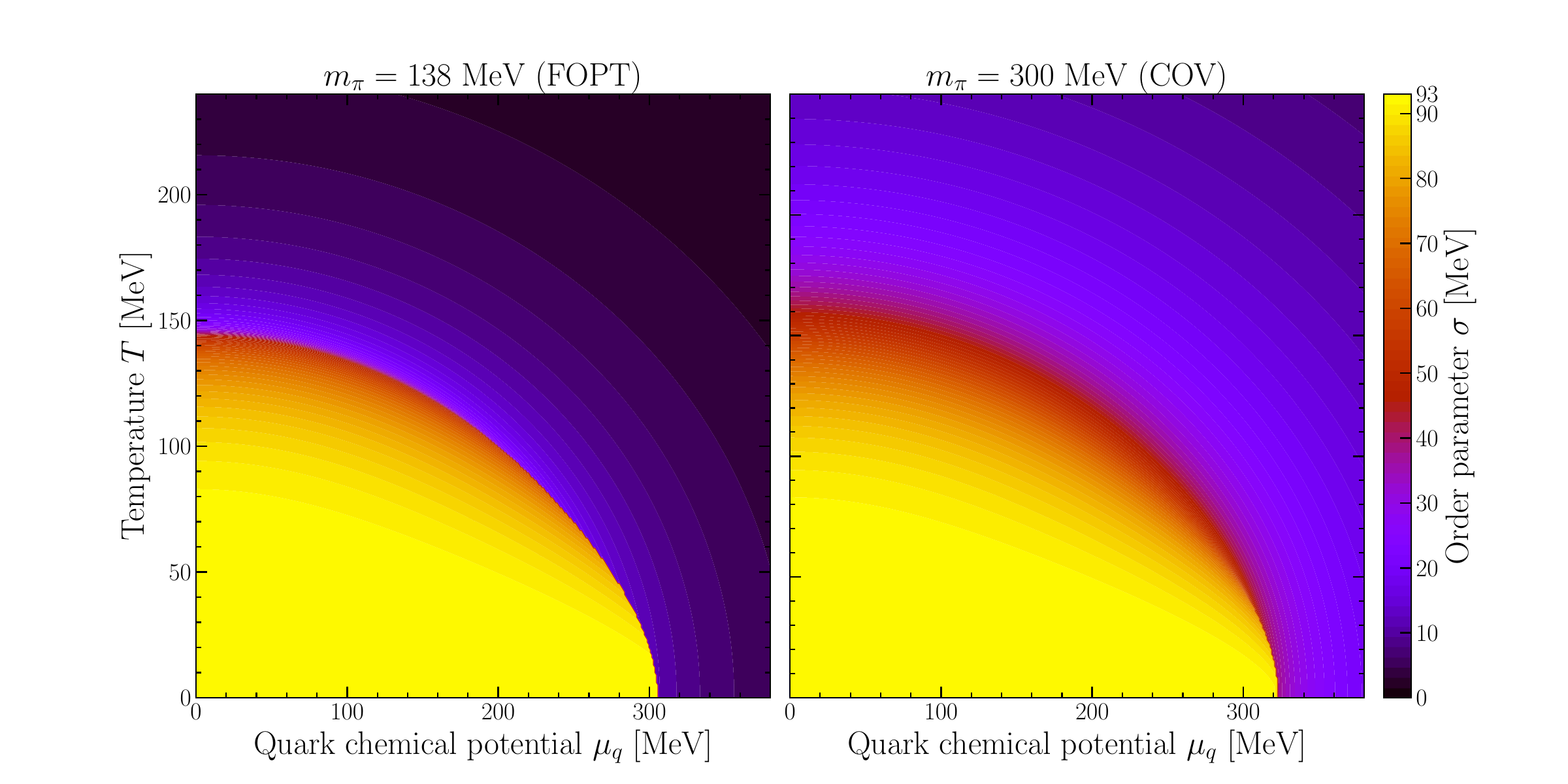}
\caption{Phase diagram in the $(\mu_q,T)$ plane for (left) $m_\pi=138~\mathrm{MeV}$ (FOPT scenario) and (right) $m_\pi=300~\mathrm{MeV}$ (COV scenario). Colors show the expectation value of the order parameter $\sigma$.}
\label{fig:phasediagram}
\end{figure*}

To study the impact of the chiral phase structure on the dynamical evolution of the medium, we use the non-equilibrium chiral fluid dynamics model (N$\chi$FD), where the phase transition emerges dynamically from the non-equilibrium evolution of the chiral order parameter $\sigma$ coupled to a Bjorken expansion of a quark fluid under dissipation and fluctuation.

The underlying effective theory is given by the quark-meson model in mean-field approximation~\cite{Nahrgang:2011mg},
\begin{align}
\mathcal{L}
&=
\bar q \left(i\gamma^\mu\partial_\mu-g\sigma\right)q
+\frac12\partial_\mu\sigma\partial^\mu\sigma
-U(\sigma)~,
\\
U(\sigma)
&=
\frac{\lambda^2}{4}\left(\sigma^2-\nu^2\right)^2-H\sigma~.
\end{align}
Here, $q=(u,d)$ denotes the light quark doublet, the explicit symmetry breaking is controlled by the linear term $H\sigma=f_\pi m_\pi^2\sigma$. The vacuum expectation value is fixed by the pion decay constant through $\nu^2=f_\pi^2-m_\pi^2/\lambda^2$. Throughout this work, the parameters are chosen as $\lambda^2=19.7$, $g=3.37$, $f_\pi=93~\mathrm{MeV}$.

The thermodynamic properties of the system are determined from the grand potential $\Omega=U+\Omega_{q\bar q}$, where the quark-antiquark contribution is given by
\begin{align}
\Omega_{q\bar q}
=
-2N_cN_fT
\int&\frac{d^3p}{(2\pi)^3}
\Bigg[
\ln\left(1+\exp\left({-\frac{E-\mu_q}{T}}\right)\right)\\\nonumber
&+\ln\left(1+\exp\left({-\frac{E+\mu_q}{T}}\right)\right)
\Bigg]~,
\end{align}
equal to the negative pressure of a gas of fermions with energies $E=\sqrt{\mathbf{p}^2+m_q^2}$. The effective (constituent) quark mass is dynamically generated by the chiral condensate,  $m_q(\sigma)=g\sigma$. The quark chemical potential is given by $\mu_q = \mu_B/3$.

The chiral order parameter $\sigma$ is propagated with a Langevin equation of motion, which for a Bjorken description reads~\cite{Herold:2018ptm}
\begin{equation}
\ddot{\sigma}
+
\left(
\frac{D}{\tau}+\eta
\right)\dot{\sigma}
+
\frac{\delta\Omega}{\delta\sigma}
=
\xi~,
\label{eq:sigma_langevin}
\end{equation}
where the damping coefficient $\eta$, which depends on $T$ and $\mu_q$~\cite{Nahrgang:2011mg}, describes dissipation into the quark heat bath and $\xi$ represents stochastic noise induced by the thermal medium. The dot represents derivatives with respect to proper time $\tau$, and we set $D=1$ in the Hubble term for a one-dimensional expansion along the beam direction.

The fluctuation-dissipation relation is implemented through~\cite{Nahrgang:2011mg}
\begin{equation}
\left\langle
\xi(\tau)\xi(\tau')
\right\rangle
=
\delta(\tau-\tau')
\frac{m_\sigma\eta}{V}
\coth\left(
\frac{m_\sigma}{2T}
\right)~,
\label{eq:fdt}
\end{equation}
ensuring thermodynamic consistency during the evolution.

The coupling between the chiral field and the fluid is realized through energy-momentum exchange,
\begin{equation}
\dot e
=
-\frac{e+P}{\tau}
+
\left[
\frac{\delta\Omega_{q\bar q}}{\delta\sigma}
+
\left(
\frac{D}{\tau}+\eta
\right)\dot\sigma
\right]\dot\sigma~,
\quad
\dot n
=
-\frac{n}{\tau}~,
\end{equation}
so that the relaxation of the order parameter feeds back into the bulk evolution. This dynamical coupling allows for supercooling, delayed phase conversion and reheating, which are characteristics of non-equilibrium effects in the vicinity of a FOPT.

For each collision energy, the initial condition is determined from the corresponding point along the Rankine-Hugoniot-Taub adiabat~\cite{Taub:1948zz,Thorne:1973} following the procedure introduced in Ref.~\cite{Bumnedpan:2022lma}. Further following this earlier work, we define the end of the evolution or freeze-out time at the point where the order parameter has passed the rapid change towards the vacuum expectation value, $d^2\sigma/d\tau^2=0$.

\subsection{Phase structure with different $m_\pi$}

The phase structure in N$\chi$FD is controlled by the explicit symmetry-breaking term of the effective chiral potential, $H=f_\pi m_\pi^2$. Changing the pion mass modifies the strength of explicit chiral symmetry breaking and reshapes the grand potential without altering the underlying dynamical framework. In the present work, this provides a systematic way to construct different phase transition scenarios while preserving the same chiral dynamics and microscopic in-medium dilepton production mechanism.

Figure~\ref{fig:phasediagram} shows the equilibrium phase diagram in the $(\mu_q,T)$ plane for the physical pion mass $m_\pi=138~\mathrm{MeV}$ (left) and for an increased pion mass $m_\pi=300~\mathrm{MeV}$ (right). The color scale represents the equilibrium expectation value of the chiral order parameter $\sigma$. Large values of $\sigma$ correspond to the chirally broken phase, small values indicate the chirally restored phase where the constituent quark mass approaches zero. While the position of the phase boundary remains largely unaltered for the two different values of $m_\pi$, the phase transition---clearly visible in the discontinuity of $\sigma$ on the left---becomes largely washed out to a smooth crossover with the increased pion mass. Furthermore, for $m_\pi=300~\mathrm{MeV}$, the critical endpoint is located at larger baryon chemical potential $(\mu_q,T)=(312,32)~\mathrm{MeV}$, compared to $(\mu_q,T)=(208,95)~\mathrm{MeV}$ in the scenario with the physical parameter. 

In the present work, the phase structure obtained with the physical pion mass is termed the FOPT scenario (left), where lower initial energies pass through the boundary of the first-order transition. The case with the increased pion mass is denoted COV since here all investigated energies evolve through the crossover region.

\subsection{Speed of sound}
\label{sec:sos}

The speed of sound provides a direct measure of the stiffness of the EoS and is therefore a sensitive indicator of the phase structure encountered during the evolution. In particular, a FOPT is expected to produce a pronounced reduction of the pressure response to compression and, under non-equilibrium conditions, may drive the system into a mechanically unstable (spinodal) region where the effective squared speed of sound becomes negative. Studying the time evolution of $c_s^2(\tau)=\partial p/\partial \varepsilon$
therefore provides a direct connection between the macroscopic dynamics of the expanding fireball and the underlying chiral phase transition. We evaluate $c_s^2$ as an effective dynamical quantity along the non-equilibrium trajectory rather than an equilibrium speed of sound. As the system approaches the coexistence region, the EoS softens and the speed of sound decreases. In the FOPT scenario, the subsequent non-equilibrium evolution drives the system into the spinodal region, where the effective squared speed of sound becomes negative. This reflects a mechanical instability of the homogeneous medium rather than merely a soft EoS. As the medium's trajectory approaches or passes the FOPT, part of the energy density is used for phase conversion rather than pressure buildup. This reduces the expansion rate and delays the relaxation of the order parameter, leading to supercooling, reheating, and an overall extended lifetime of the dense medium.

\begin{figure}[hbt!]
    \centering
    \includegraphics[width=\columnwidth]{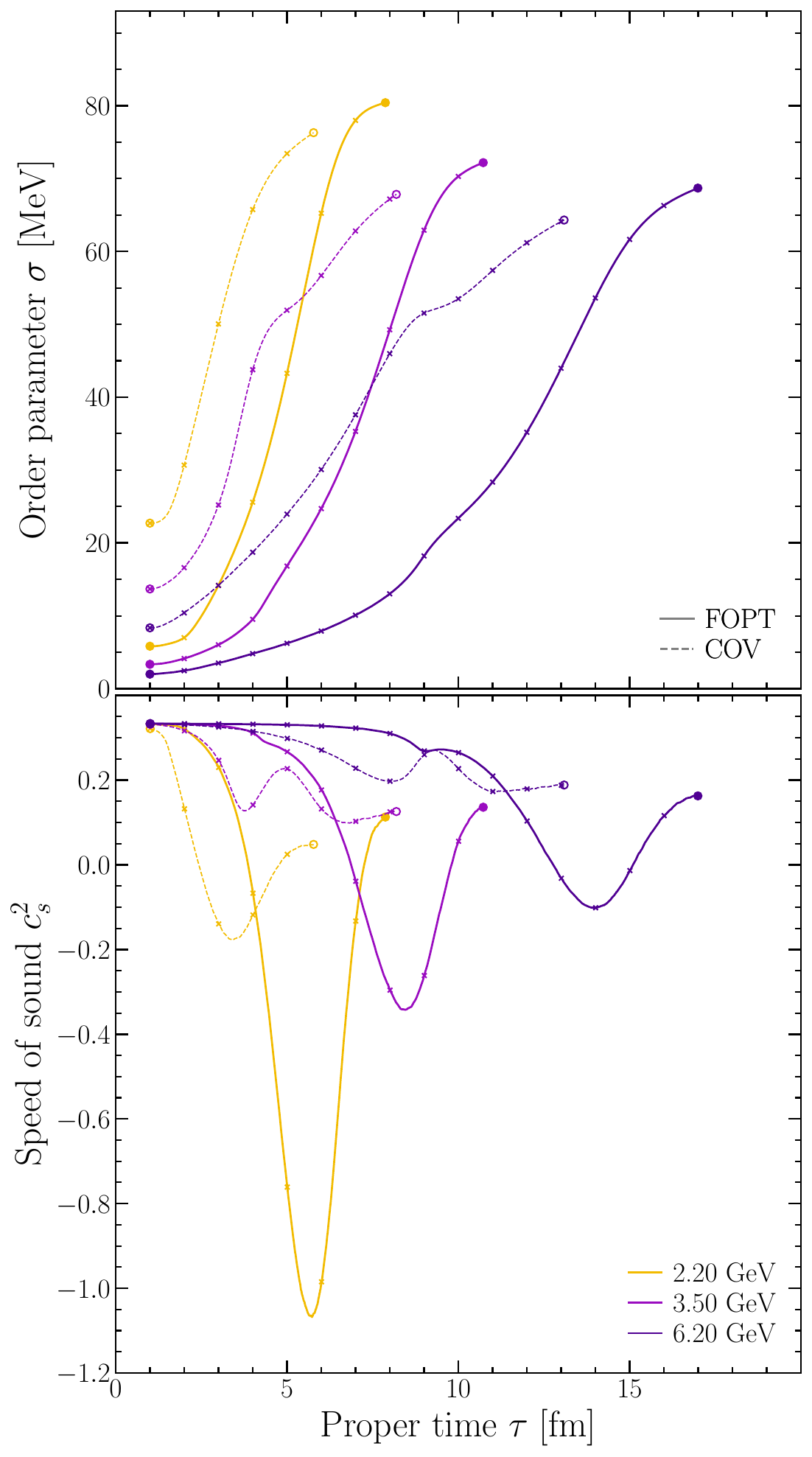}
    \caption{The evolution of the order parameter in proper time (top) and of the squared speed of sound $c_s^2(\tau)$ (bottom) for FOPT (solid) and COV (dashed) scenarios at different center-of-mass energies.}
    \label{fig:freezeout}
\end{figure}

Figure~\ref{fig:freezeout} shows the time evolution of the order parameter $\sigma(\tau)$ (top) and the squared speed of sound $c_s^2(\tau)$ (bottom) at three different center-of-mass energies. Solid lines correspond to the FOPT scenario, while dashed lines represent the COV scenario. The crosses along the lines indicate proper time intervals of $1~\mathrm{fm}/c$ for easier comparison with the trajectories in figure~\ref{fig:traj}. For all energies, the order parameter $\sigma$ remains near the chirally restored value for a longer period in the FOPT compared to the COV scenario, before rapidly recovering towards its vacuum value. 

All systems start near $c_s^2\simeq1/3$, characteristic of a stiff relativistic medium. During expansion, $c_s^2$ decreases for each energy in both scenarios, reflecting the softening of the EoS. In the FOPT scenario, $c_s^2$ becomes strongly negative, indicating that the system enters a mechanically unstable (spinodal) region associated with the non-equilibrium FOPT. The deepest negative values occur at $\sqrt{s_{\rm NN}}=2.20$~GeV and become progressively less pronounced with increasing energy. In contrast, the COV scenario exhibits only a modest reduction of $c_s^2$, which evolves close to or slightly below zero.

The correlation between the development of a pronounced negative $c_s^2$ and the delayed recovery of $\sigma$ indicates a stronger coupling between the thermodynamic evolution and the chiral dynamics in the FOPT scenario. This additional delay ultimately leads to a later freeze-out relative to the corresponding COV evolution.

\subsection{Non-equilibrium trajectories}

\begin{figure}[t]
\centering
\includegraphics[width=\columnwidth]{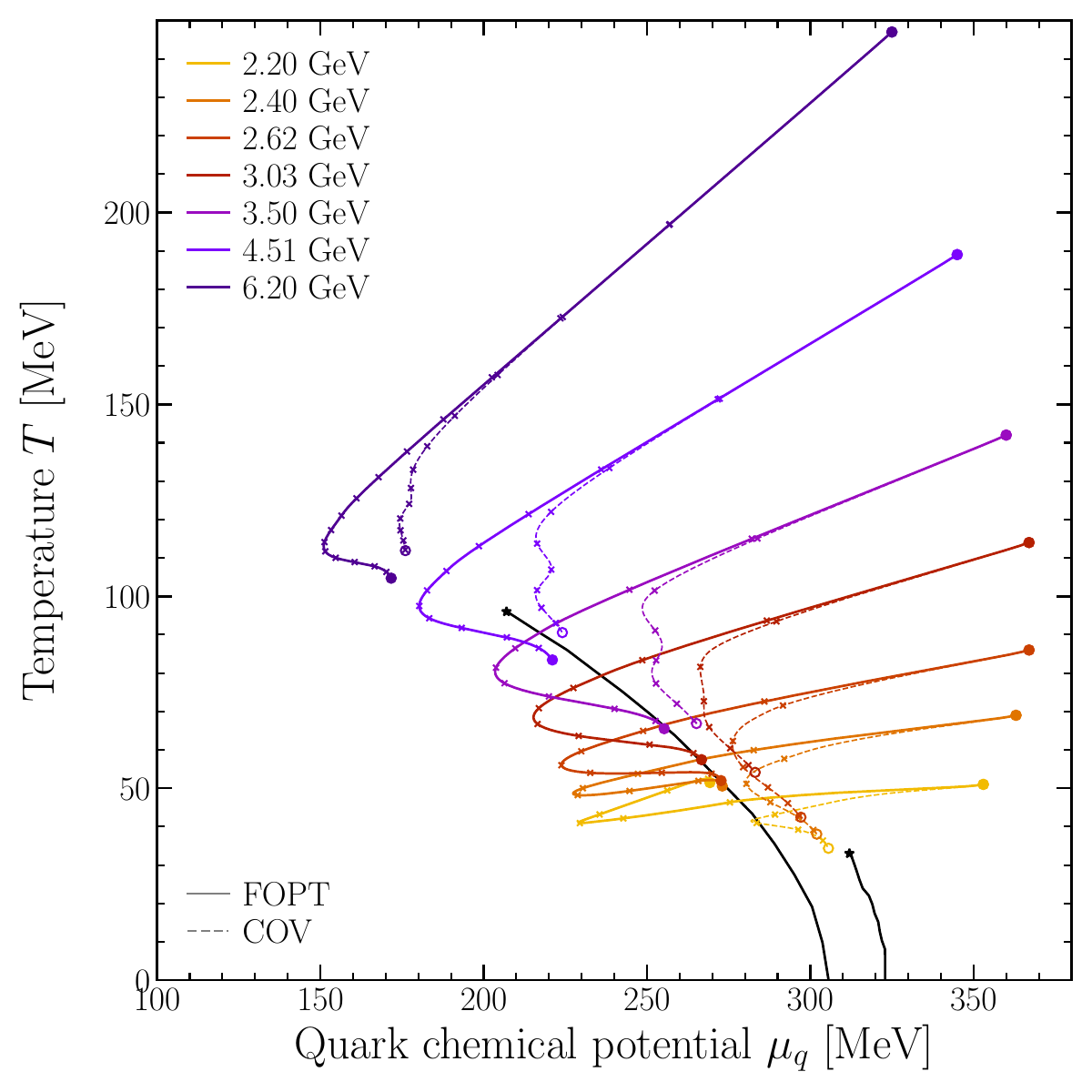}
\caption{Non-equilibrium trajectories in the $(\mu_q,T)$ plane for different center-of-mass energies. Solid lines correspond to the FOPT scenario and dashed lines to the COV scenario. The black solid lines delineate the phase transition boundaries in each EoS and the star symbols mark the positions of the respective critical endpoint.}
\label{fig:traj}
\end{figure}

Figure~\ref{fig:traj} shows the non-equilibrium trajectories in the $(\mu_q,T)$ plane for the FOPT (solid lines) and COV (dashed lines) scenarios. The equilibrium FOPT boundaries and critical endpoints are included to illustrate how the dynamical trajectories probe different regions of the phase diagram.

For the FOPT scenario, the trajectories at $\sqrt{s_{\rm NN}}=2.20$--$3.50~\mathrm{GeV}$ cross the first-order phase boundary, whereas the trajectories at $\sqrt{s_{\rm NN}}=4.51$ and $6.20~\mathrm{GeV}$ remain within the crossover region. In contrast, all trajectories in the COV scenario evolve through the crossover. These different trajectory classes correlate with the beam-energy dependence of the mechanical instability and delayed relaxation discussed in the previous section~\ref{sec:sos}.

The strongest non-equilibrium effects occur for the $\sqrt{s_{NN}}=2.20~\mathrm{GeV}$ FOPT trajectory, where the system penetrates deepest into the first-order region. Here, the order parameter substantially lags behind its equilibrium value, producing pronounced supercooling followed by reheating. The associated release of latent heat generates the characteristic back-bending of the trajectory and coincides with the extended interval of negative $c_s^2$ in figure \ref{fig:freezeout}. The same is observed at $2.40-3.50$~GeV, although the effect decreases with increasing beam energy.

The dynamical response is governed not only by whether a trajectory crosses the first-order boundary, but also by its proximity to the critical region. Although the $\sqrt{s_{NN}}=4.51$ and $6.20~\mathrm{GeV}$ trajectories in the FOPT scenario remain within the crossover regime, their close approach to the first-order boundary still produces a noticeable suppression of $c_s^2$ together with a relatively rapid recovery of the order parameter, resembling the response of the lower-energy first-order trajectories. A similar behavior is observed for $\sqrt{s_{NN}}=2.20~\mathrm{GeV}$ in the COV scenario, which evolves close to the corresponding first-order boundary despite remaining on the crossover side. As the trajectories move farther from the critical region, the thermodynamic response becomes progressively weaker.

These different trajectories determine the time dependence of the temperature and baryon chemical potential entering the dilepton emission rates. Relative to the COV scenario, the FOPT trajectories remain longer in the non-equilibrium regime before phase conversion, modifying both the duration and the thermodynamic conditions of dilepton production. Consequently, the largest differences in the dilepton spectra are expected for trajectories that cross or pass closest to the first-order phase boundary.

\section{In-medium vector-meson properties from resonance-driven forward scattering}
\label{sec:selfenergy}

To isolate the effects of the EoS from purely hadronic in-medium modifications, the dilepton emission rates are calculated using identical microscopic vector-meson self-energies for both EoS scenarios. The in-medium properties of the $\rho$ and $\omega$ mesons are evaluated using the forward scattering framework of Eletsky \textit{et al.}~\cite{Eletsky:2001bb}. In this approach, the retarded vector-meson self-energy is constructed from hadronic scattering amplitudes with pions and nucleons.

Compared to the original implementation of Ref.~\cite{Eletsky:2001bb}, the present work extends the resonance content using the full hadronic PDG resonance table employed in UrQMD (Table~\ref{tab:merged_urqmd_eletsky}). These resonances enter through their masses, widths, branching ratios, and quantum numbers, while the macroscopic evolution is provided by N$\chi$FD.

\subsection{Forward scattering amplitude and kinematics}

In the $V-a$ center-of-mass frame, where $V=\rho,\omega$ and $a=\pi,N$, the forward scattering amplitude is written as the sum of Breit-Wigner resonance contributions and a non-resonant background,
\begin{equation}
f_{\mathrm{c.m.}}^{Va}(s)
=
f_{\mathrm{c.m.}}^{Va,\mathrm{BW}}(s)
+
f_{\mathrm{c.m.}}^{Va,\mathrm{bg}}(s)~,
\label{eq:fcm_total}
\end{equation}
with the Mandelstam variable $s$ and the subscript ``c.m.'' referring to the center-of-mass frame. The center-of-mass momentum is given by
\begin{equation}
q_{\mathrm{c.m.}}(s)=\frac{1}{2\sqrt{s}}
\sqrt{\left[s-(m_V+m_a)^2\right]\left[s-(m_V-m_a)^2\right]}~, 
\label{eq:qcm_complete}
\end{equation}
and the kinetic energy is evaluated as
\begin{equation}
E_V-m_V=\frac{s-(m_V+m_a)^2}{2m_a}~,
\label{eq:Em_complete}
\end{equation}
which vanishes at threshold and is convenient for representing the threshold structure of $\operatorname{Im}f_{\mathrm{c.m.}}^{Va}$.

At low center-of-mass energies, the scattering is dominated by $s$-channel resonance formation. The resonance contribution is constructed as a sum over intermediate Breit-Wigner resonances $R$,
\begin{equation}
f_{\mathrm{c.m.}}^{Va,\mathrm{BW}}(s)
=
\frac{1}{2q_{\mathrm{c.m.}}(s)}
\sum_{R}
W_R^{Va}\,
\frac{\Gamma_R(s)\,\mathrm{BR}_{R\rightarrow Va}}
{M_R-\sqrt{s}-\frac{i}{2}\Gamma_R(s)}~,
\label{eq:BW_complete}
\end{equation}
where $M_R$ and $\Gamma_R$ denote the resonance mass and width, respectively, and $\mathrm{BR}_{R\rightarrow Va}$ is the branching ratio into the corresponding channel. The statistical spin--isospin factor is
\begin{equation}
W_R^{Va}
=
\frac{(2J_R+1)}{(2J_V+1)(2J_a+1)}
\frac{(2I_R+1)}{(2I_V+1)(2I_a+1)}~.
\label{eq:W_complete}
\end{equation}
The energy dependence of $\Gamma_R(s)$ controls the threshold behavior of $\operatorname{Im}f_{\mathrm{c.m.}}^{Va}$.

The non-resonant contributions in equation~\eqref{eq:fcm_total} are described by a Pomeron term,
\begin{equation}
f_{\mathrm{c.m.}}^{Va,\mathrm{bg}}(s)
=
-\frac{q_\mathrm{c.m.}(s)}{4\pi s}
\frac{1+\exp(-i\pi\alpha_P)}{\sin(\pi\alpha_P)}\,
r^{Va}_P s^{\alpha_P}~.
\label{eq:Regge_complete}
\end{equation}
The parameters are adopted from Ref.~\cite{Eletsky:2001bb} as $\alpha_P=1.093$ with corresponding residues $r_P^{V\pi}=7.508$, $r_P^{VN}=11.88$.

\begin{figure}[hbt!]
\centering
\includegraphics[width=0.49\textwidth]{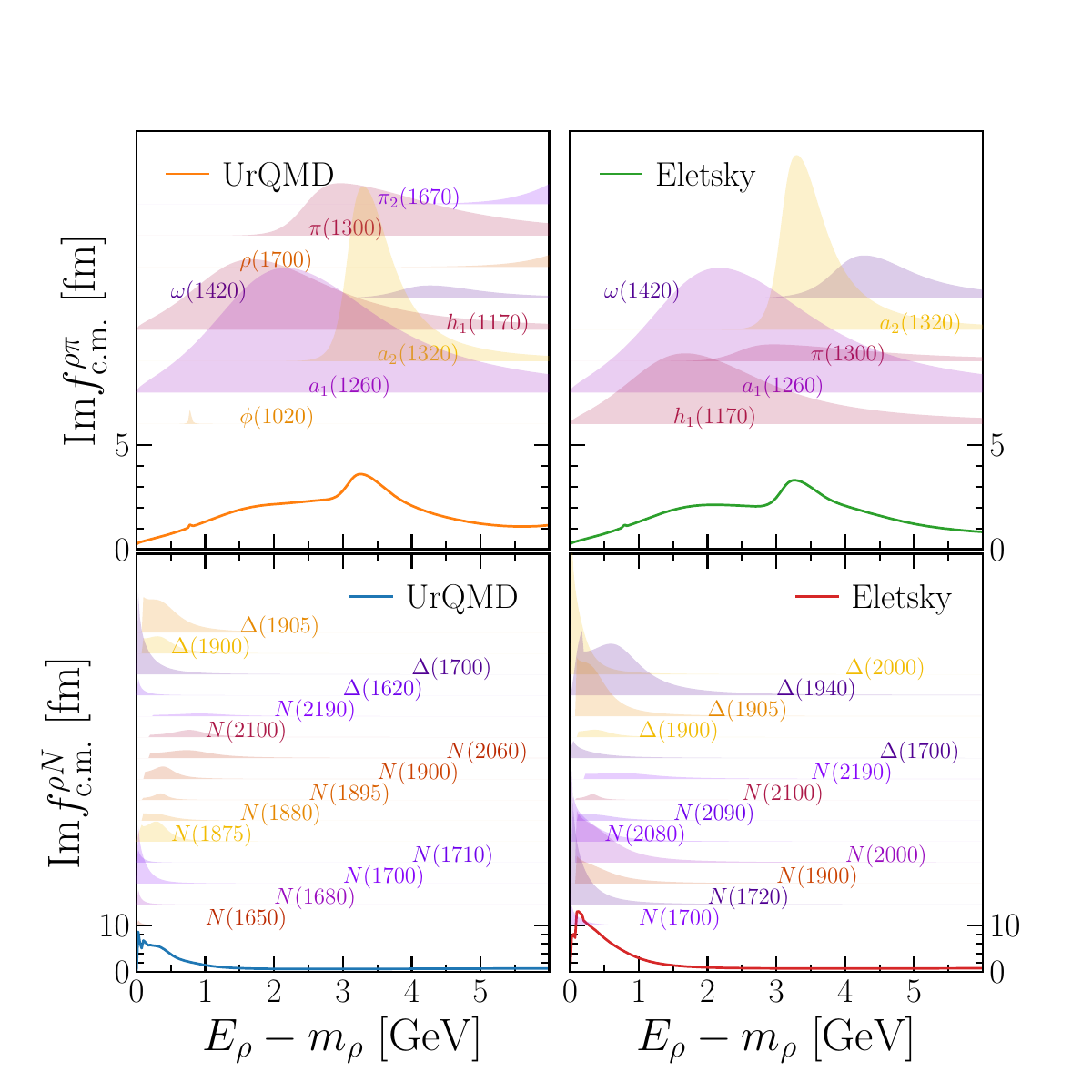}
\includegraphics[width=0.49\textwidth]{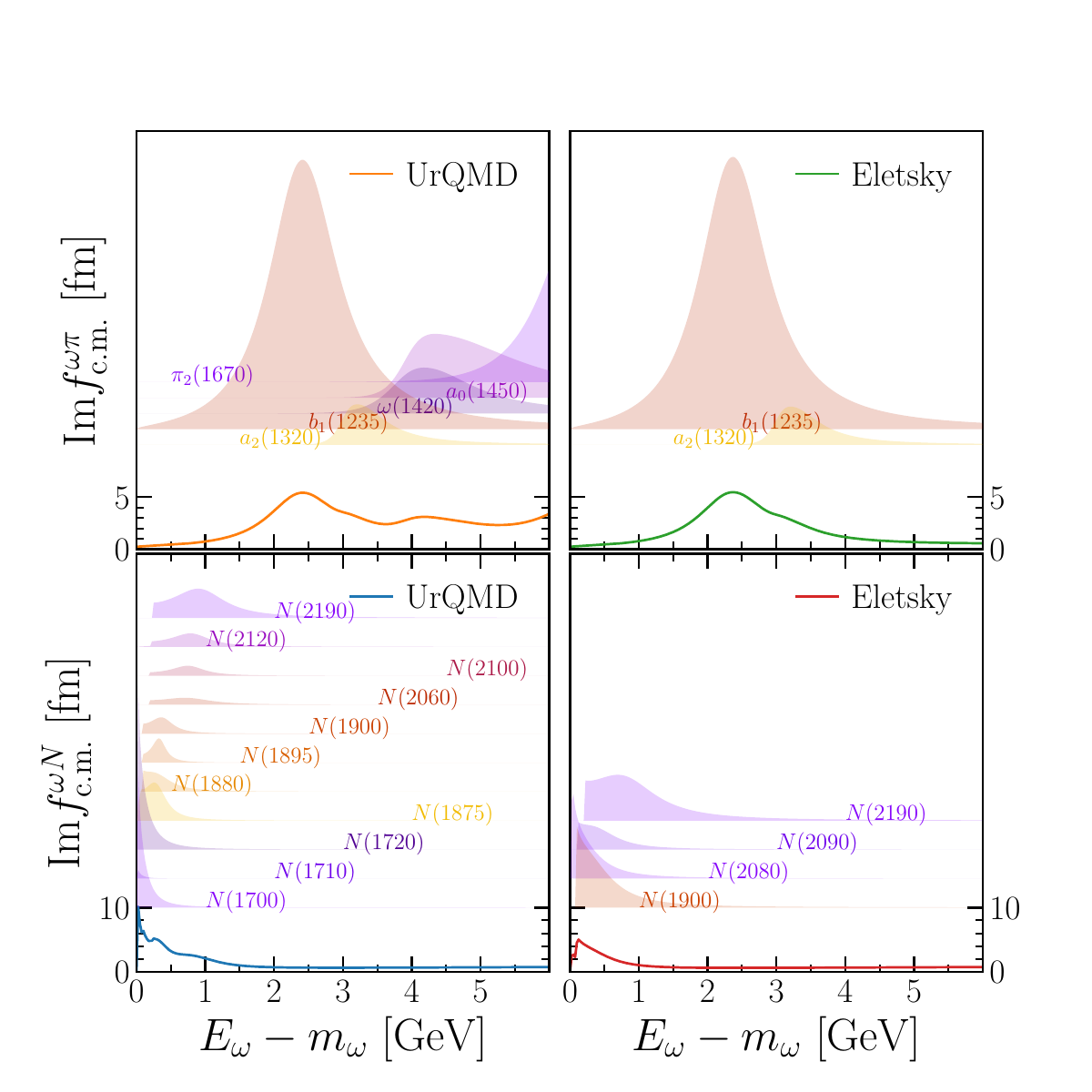}
\caption{Contributions of individual resonances to $\operatorname{Im}f^{Va}_\mathrm{c.m.}$ as a function of $E_V-m_V$. From top to bottom, the rows correspond to the $\rho\pi$, $\rho N$, $\omega\pi$, and $\omega N$ scattering channels, while the left and right columns compare the extended UrQMD/PDG resonance set with the original Eletsky \textit{et al.} implementation \cite{Eletsky:2001bb}. Colored fills illustrate individual resonance contributions, and the solid lines represent the total $\operatorname{Im}f^{Va}_\mathrm{c.m.}$.}
\label{fig:Imf}
\end{figure}

The optical theorem relates the imaginary part of the forward scattering amplitude to the total cross section via $\sigma^{Va}=4\pi \operatorname{Im}f_{\mathrm{c.m.}}^{Va}/q_{\mathrm{c.m.}}$. Figure~\ref{fig:Imf} compares $\operatorname{Im}f_{\mathrm{c.m.}}^{Va}$ obtained from the extended UrQMD/PDG resonance set against the original implementation of Ref.~\cite{Eletsky:2001bb}. The two approaches agree closely for the $\rho\pi$, $\rho N$, and $\omega N$ channels, indicating that the dominant contributions to the $\rho$ self-energy are already included in the original resonance set. The largest difference appears in the $\omega\pi$ channel, where the additional resonances enhance the scattering amplitude at higher kinetic energies and consequently increase the in-medium broadening of the $\omega$ meson.

\subsection{Retarded self-energy}

The contribution from hadronic scattering to the retarded self-energy of a vector meson $V$ with four-momentum $p=(E,\mathbf{p})$ is obtained from the forward scattering amplitude,
\begin{align}
\label{eq:Sigma_complete}
\Sigma_V^{R}(p)
=-4\pi \sum_{a}\int &\frac{\mathrm{d}^3 \mathbf{k}}{(2\pi)^3}
n_a(\omega_k;T,\mu_B)\frac{\sqrt{s}}{\omega_k}\,
f^{Va}_\mathrm{c.m.}(s)~.
\end{align}
Here,  $\omega_k=\sqrt{\mathbf{k}^2+m_a^2}$,  $s=(p+k)^2$, and $n_a$ is the Bose--Einstein or Fermi--Dirac distribution.

For the vacuum contribution, only the $\rho$ self-energy is explicitly included. The vacuum self-energy of the $\omega$ meson is not considered since its dominant decay channel, $\omega\rightarrow3\pi$, proceeds through a three-body hadronic final state and therefore lies outside the two-body forward-scattering framework employed in the present work. Furthermore, the vacuum width of $\omega$ remains considerably smaller than that of $\rho$, so its omission has only a minor impact on the resulting spectral function. The vacuum $\rho$ self-energy is implemented as

\begin{align*}
    \operatorname{Re}\Sigma_\rho^\mathrm{vac}=&\frac{g_\rho^2M^2}{48\pi^2}\left[\left(1-\frac{4m_\pi^2}{M^2}\right)^{3/2}\operatorname{ln}\left| \frac{1+\sqrt{1-\frac{4m_\pi^2}{M^2}}}{1-\sqrt{1-\frac{4m_\pi^2}{M^2}}} \right| \right.\\
    &+ 8m_\pi^2\left(\frac{1}{M^2}-\frac{1}{m_\rho^2}\right) - \left. 2\left( \frac{p_0}{\omega_0}\right)^3 \operatorname{ln}\frac{\omega_0 + p_0}{m_\pi}\right]~,
\end{align*}
\begin{align}
    \operatorname{Im}\Sigma_V^\mathrm{vac} = -\frac{g_\rho^2M^2}{48\pi}\left( 1 -\frac{4m_\pi^2}{M^2}\right)^{3/2}~,
\end{align}
where $\omega_0=\sqrt{m_\pi^2+p_0^2}$, and $p_0$ is evaluated from the vacuum width $\Gamma_\rho^\mathrm{vac}=\frac{g_\rho^2}{48\pi}m_\rho\left(\frac{p_0}{\omega_0}\right)^3$. The coupling constants of vector mesons to photons are given by $g_\rho^2 / 4\pi=2.54$ and $g_\omega^2=8g_\rho^2$.

\subsection{Propagator, spectral function, and dilepton rate}

The propagation of a vector meson inside hot and dense matter is modified by its interactions with the surrounding medium through the total self-energy $\Sigma_V\equiv \Sigma_V^R + \Sigma_V^\mathrm{vac}$. The corresponding in-medium retarded propagator is written as
\begin{equation}
D_V^{R}(p) =
\frac{1}{E^2-\mathbf{p}^2-m_V^2-\Sigma_V(p)}~.
\label{eq:DV}
\end{equation}
The imaginary part of the propagator determines the in-medium spectral function and therefore encodes how the vector meson is redistributed in invariant mass due to rescattering. The thermal dilepton emission rate is expressed through the imaginary part of the electromagnetic current correlator~\cite{Rapp:2013nxa},
\begin{align}
\label{eq:rate_master}
\frac{\mathrm d N}{\mathrm d^4x\,\mathrm d^4p}
=
-&\frac{\alpha^2}{\pi^3 M^2}\left(1+\frac{2m^2_\ell}{M^2}\right)\sqrt{1-\frac{4m^2_\ell}{M^2}}\\\nonumber
&\times n_B(E;T)
\operatorname{Im}\Pi_{\mathrm{em}}^{R}(M,p;T,\mu_q)~,
\end{align}
with fine structure constant $\alpha$, lepton mass $m_\ell$, and invariant mass of the dilepton $M$.
This correlator is constructed from the in-medium $\rho$ and $\omega$ propagators~\cite{vanHees:2007th},
\begin{equation}
\operatorname{Im}\Pi_{\mathrm{em}}^{R}(p)
=
\sum_{V=\rho,\omega}
\left(\frac{m_V^2}{g_V}\right)^2
\operatorname{Im}D_V^{R}(p)~.
\label{eq:imPiem}
\end{equation}

The resonance-driven in-medium scattering amplitudes directly determine the dilepton emissivity along the evolving non-equilibrium trajectories. Since the dilepton rate depends on the in-medium spectral distribution, modifications of the vector-meson propagator induced by the dense hadronic environment become observable through the invariant mass spectra of dileptons.

The medium evolution is modeled by a longitudinal Bjorken expansion with fixed transverse radius $R_T=6.5$~fm to resemble the longitudinal expansion of a central Au+Au collision. While this simplified geometry affects the overall normalization of the dilepton yield, both EoS scenarios are treated identically so that the relative FOPT–COV comparison remains unaffected.

\section{Time evolution of mass integrated dilepton yields}
\label{sec:timeprofile}

\begin{figure}[t]
\centering
\includegraphics[width=0.49\textwidth]{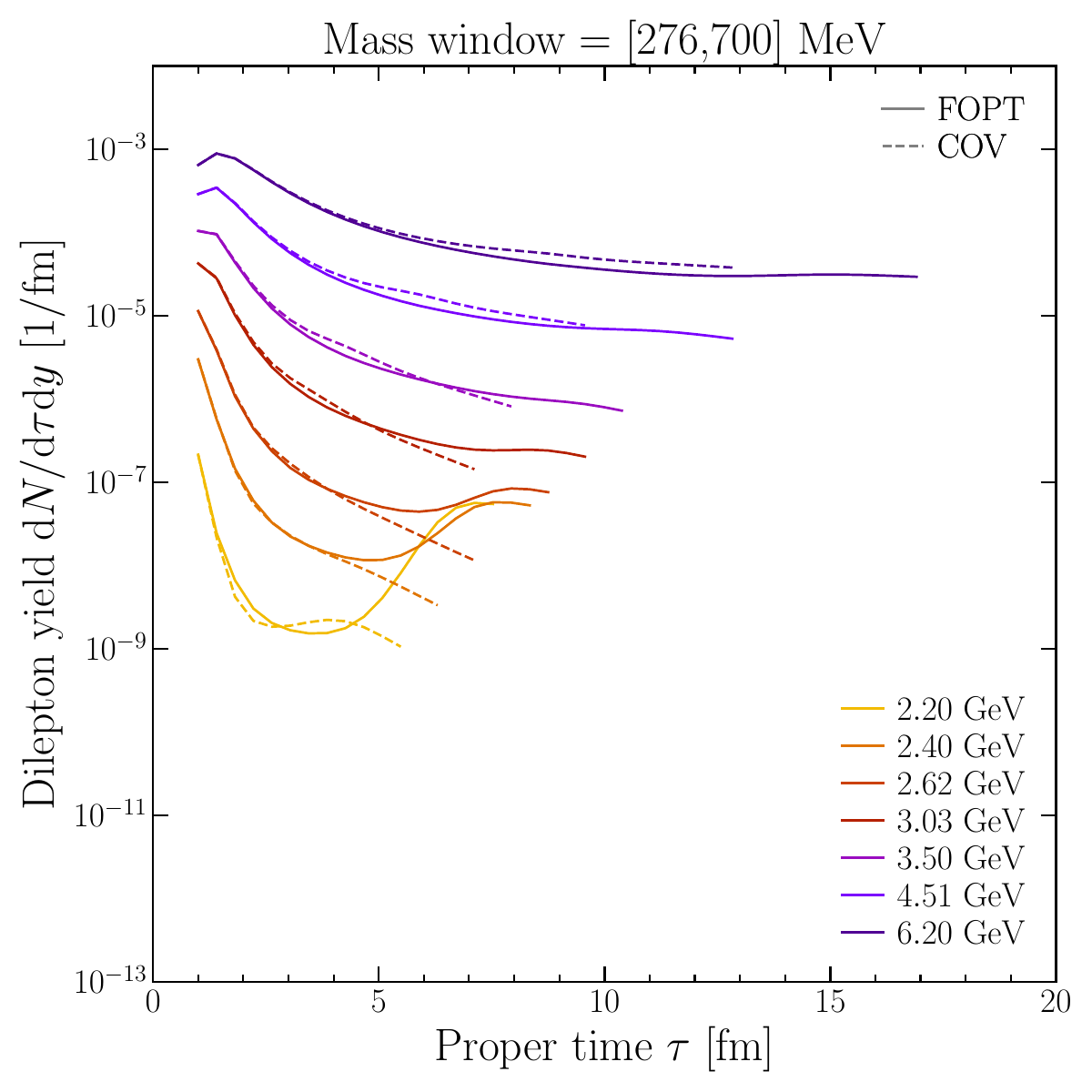}
\includegraphics[width=0.49\textwidth]{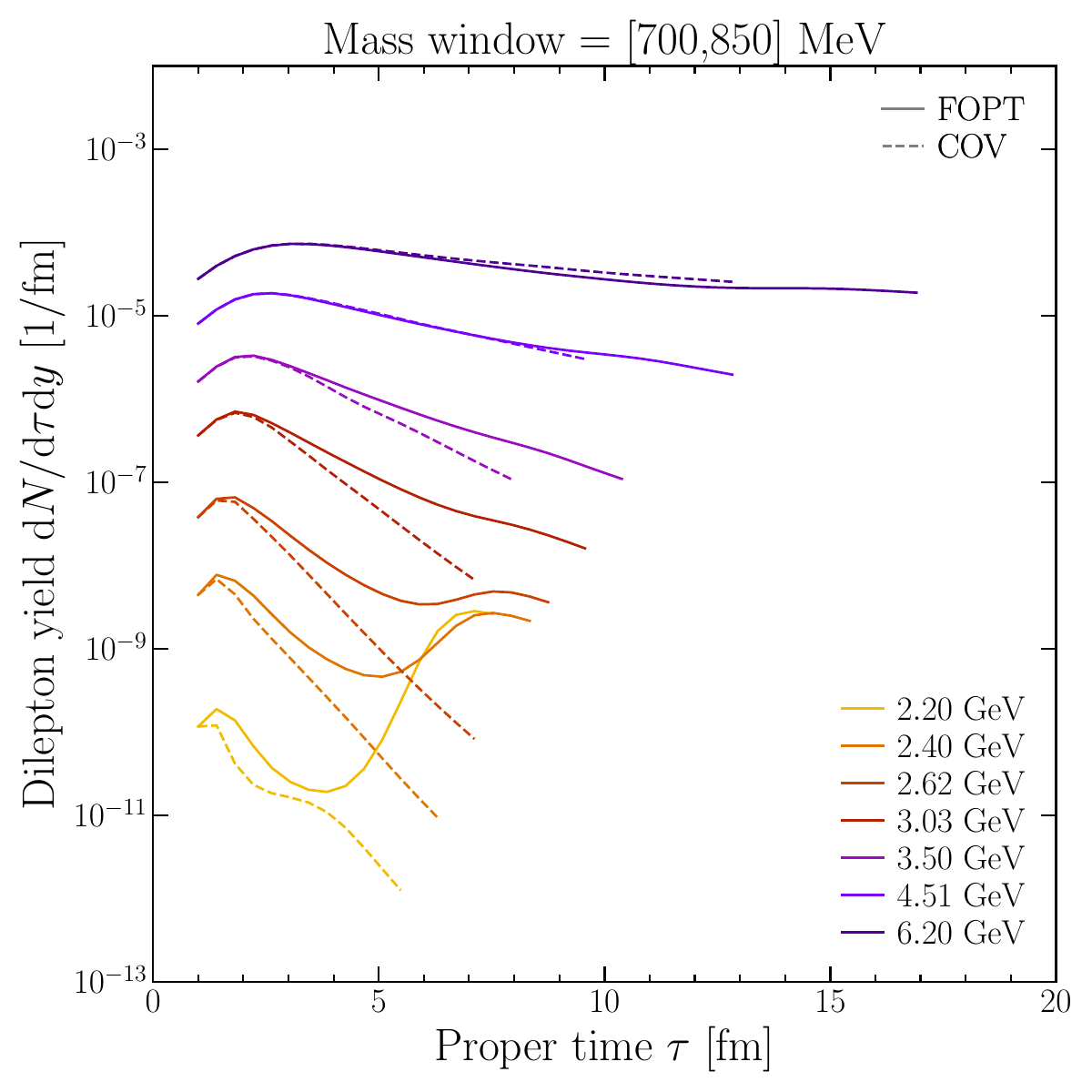}
\caption{Time profiles of $\mathrm{d}N/\mathrm{d}\tau\,\mathrm{d}y$ at mid-rapidity integrated over (top) low-mass $M\in[276,700]~\mathrm{MeV}$ and (bottom) pole-mass $M\in[700,850]~\mathrm{MeV}$.}
\vspace{0.4em}
\label{fig:dNdtaudy_AB}
\end{figure}

We examine the time dependence of the emission rate before integrating over the full fireball evolution. This allows the different stages of the collision to be separated and reveals whether the low- and pole-mass regions respond differently to the underlying phase structure.

We integrate the in-medium dilepton emission rate in equation~\eqref{eq:rate_master} over two selected invariant mass windows at midrapidity to obtain the corresponding time-resolved rates. The two mass intervals are chosen to separate different spectral sensitivities: 
\begin{itemize}
\item[\textbullet] a low-mass window $M\in[276,700]~\mathrm{MeV}$,
\item[\textbullet] a pole-mass window $M\in[700,850]~\mathrm{MeV}$, dominated by the $\rho$ and $\omega$ region.
\end{itemize}


Figure~\ref{fig:dNdtaudy_AB} shows the rate for both mass windows and reveals the central result of this section: the FOPT produces two temporally distinct dilepton signatures. The pole-mass region $[700,850]$~MeV responds already during the early evolution, whereas the low-mass enhancement develops predominantly during the later reheating stage.

As expected from the analysis of the trajectories, the largest differences between FOPT and COV occur for $\sqrt{s_{\rm NN}}=2.20-3.50$~GeV. At higher beam energies, both scenarios evolve through the crossover region and therefore exhibit similar emission histories.

The low-mass enhancement  for $M\in[276,700]~\mathrm{MeV}$ is primarily a consequence of the overshooting and back-bending of the FOPT trajectories and the visible reheating at the three lowest energies in figure~\ref{fig:traj}. The recovery of temperature increases the thermal dilepton emission, while the simultaneous recovery of baryon density strengthens the resonance-driven vector-meson self-energies and redistributes additional spectral strength into the low-mass continuum. Consequently, the low-mass window is dominated by accumulated late-stage emission rather than by the initial response to the onset of EoS softening.

In contrast, the pole-mass enhancement for $M\in[700,850]~\mathrm{MeV}$ originates during the supercooled stage of the FOPT and is clearly visible for $\sqrt{s_{\rm NN}}=2.20-3.50$~GeV, getting significantly more pronounced with decreasing energy. Here, the FOPT trajectories exhibit baryon densities lower than those of the corresponding COV trajectories, reducing the baryonic contribution to the vector-meson self-energies, while the comparatively high temperature further enhances the thermal emission rate. Together, these effects contribute to the observed pole-mass enhancement.

The time-resolved dilepton emission demonstrates that the FOPT does not simply increase the total yield but modifies when different invariant mass regions are emitted. The pole-mass region probes the early non-equilibrium evolution associated with the onset of the phase transition, whereas the low-mass continuum is dominated by the prolonged reheating stage. These distinct temporal signatures provide the physical origin of the differences observed later in the integrated invariant mass spectra. Although the emission time itself is not experimentally accessible, these distinct production stages leave characteristic imprints on different invariant mass regions after integration over the full evolution.

\section{Dilepton spectra}
\label{sec:rate}

\begin{figure}[t]
\centering
\includegraphics[width=\columnwidth]{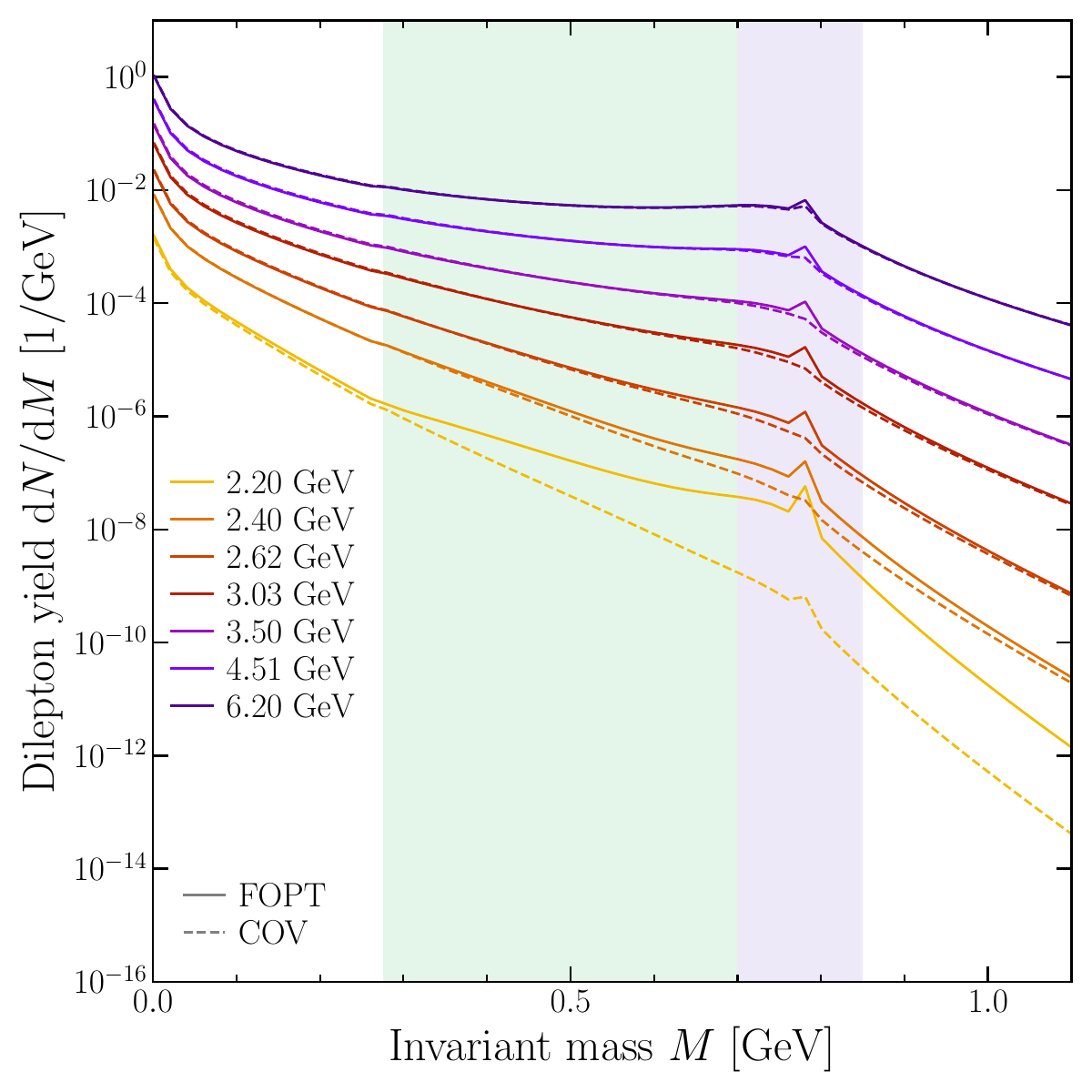}
\caption{Invariant mass spectra $\mathrm{d}N/\mathrm{d}M$ for  FOPT (solid lines) and COV (dashed lines) scenarios. Shaded bands indicate the mass windows $[276,700]~\mathrm{MeV}$ (green) and $[700,850]~\mathrm{MeV}$ (purple).}
\label{fig:urqmd_all}
\end{figure}

\begin{figure}[ht]
\centering
\includegraphics[width=\columnwidth]{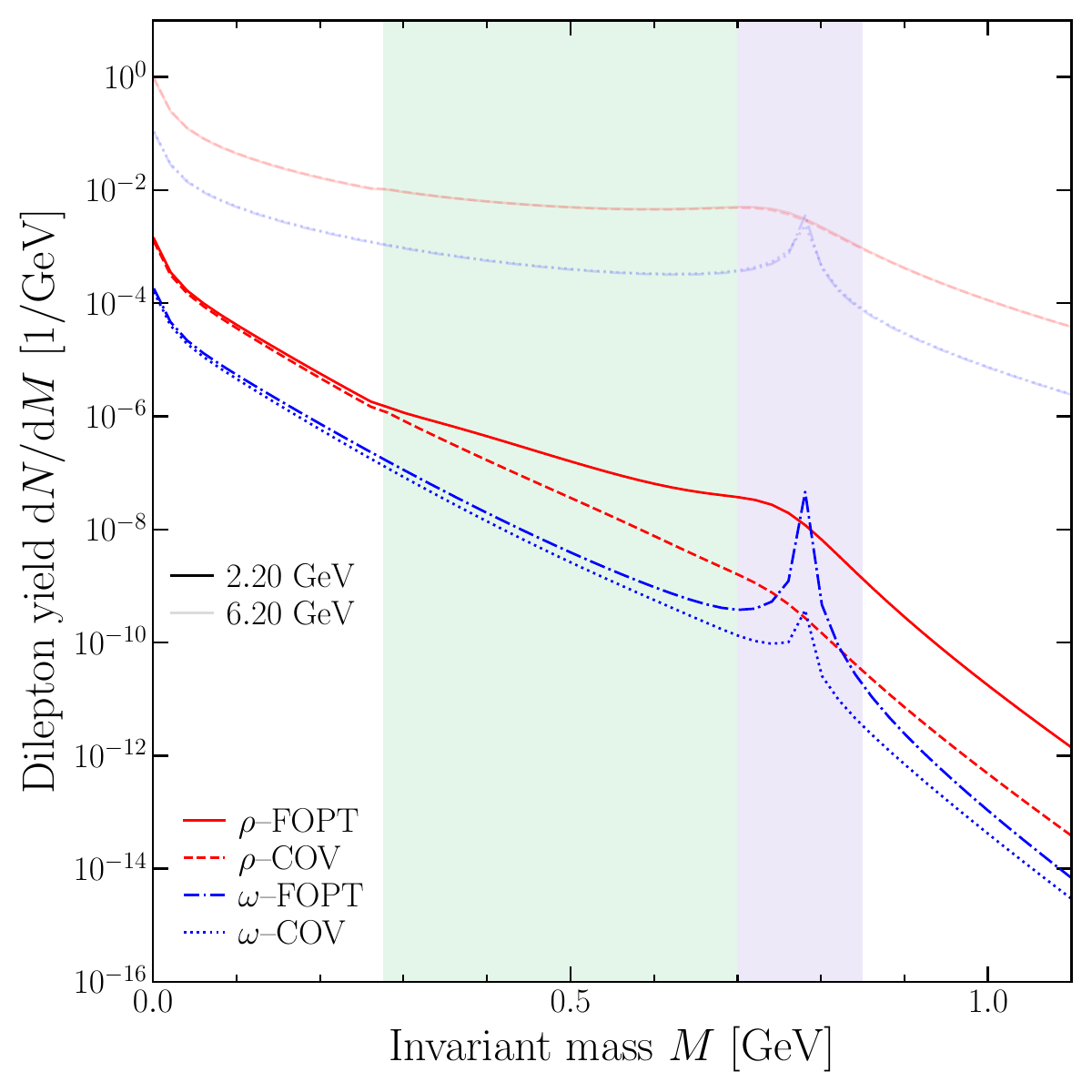}
\caption{Separated invariant mass spectra $\mathrm{d}N/\mathrm{d}M$ of dileptons for two center-of-mass energies: $\rho$ contribution (red) and $\omega$ contribution (blue). Shaded bands indicate the mass windows $[276,700]~\mathrm{MeV}$ (green) and $[700,850]~\mathrm{MeV}$ (purple).}
\label{fig:urqmd_sep}
\end{figure}

We now investigate how the temporally distinct emission patterns identified in the previous section are encoded in the experimentally observable invariant mass spectrum. Integrating over the full fireball evolution removes the explicit time information, so any remaining differences between the FOPT and COV scenarios must arise from the cumulative effect of their different dynamical histories.

Figure~\ref{fig:urqmd_all} presents the total dilepton invariant mass spectra for all energies, while figure~\ref{fig:urqmd_sep} separates the individual $\rho$ (red) and $\omega$ (blue) contributions for the representative lowest and highest investigated energies $\sqrt{s_{NN}}=2.20$ and $6.20~\mathrm{GeV}$. Solid lines correspond to the FOPT scenario and dashed lines to the COV scenario. The shaded bands indicate the low-mass and pole-mass windows introduced in the time-profile analysis in the previous section.

The overall energy dependence is driven by the combined increase in temperature, lifetime, emission volume, and in-medium broadening of the vector mesons. Similarly, the spectral strength is progressively redistributed from the vector-meson pole region toward lower invariant masses. The hotter and denser medium enhances the in-medium broadening of the vector-meson spectral functions through resonance-driven hadronic scattering, producing a stronger low-mass continuum and less pronounced pole structures at higher center-of-mass energies.

Superimposed on this common energy dependence is the characteristic signature of the FOPT. At $\sqrt{s_{NN}}=2.20$--$3.50~\mathrm{GeV}$, where the trajectories cross the first-order boundary, the FOPT scenario exhibits a visible excess over the corresponding COV spectra in both invariant mass regions, see figure~\ref{fig:urqmd_all}. The largest FOPT--COV separation is observed at $2.20~\mathrm{GeV}$ and is visible in both the low-mass and pole-mass regions. As demonstrated by the time-resolved emission profiles in figure~\ref{fig:dNdtaudy_AB}, the additional pole-mass yield originates predominantly during the early non-equilibrium evolution, whereas the low-mass enhancement accumulates during the reheating stage. Integration over the full evolution preserves both contributions, producing the enhanced FOPT spectra observed here.

Figure~\ref{fig:urqmd_sep} sheds light on how the microscopic origin of the spectral modification differs between the two vector mesons. Due to its strong coupling to the $\pi\pi$ channel, the $\rho$ meson already possesses a broad vacuum spectral function. Resonance-driven $\rho N$ and $\rho\pi$ scattering further enhance the in-medium self-energy $\operatorname{Im}\Sigma_\rho^R$, transferring spectral strength from the pole region into the low-mass continuum. Consequently, the $\rho$ spectrum exhibits substantial broadening together with a pronounced enhancement of the low-mass tail.

The $\omega$ meson exhibits the same qualitative in-medium response through resonance-driven scattering with surrounding hadrons. However, because of its much narrower width, the medium-induced broadening remains considerably weaker than that of the $\rho$, allowing the pole structure to remain clearly visible after integration over the full evolution. Consequently, the FOPT enhancement manifests primarily as an increased pole yield rather than a substantial redistribution of spectral strength.

The invariant mass spectra therefore demonstrate that the dynamical effects of a FOPT survive the integration over the complete fireball evolution. Although the temporal information is no longer directly accessible, the distinct early and late emission mechanisms remain encoded in different invariant mass regions. Because the microscopic vector-meson self-energies are identical in the FOPT and COV calculations, the observed spectral differences can be attributed directly to the different macroscopic evolution induced by the EoS.

\section{Excitation function of dilepton yield }
\label{sec:excitation}

The excitation function of the total dilepton yield in the two selected mass windows provides another experimentally accessible observable of the impact of a phase transition on dilepton production. Unlike the time-dependent emission rates or invariant mass spectra, it directly reveals how the sensitivity to the EoS evolves with initial energy.

\begin{figure}[t]
\centering
\includegraphics[width=0.49\textwidth]{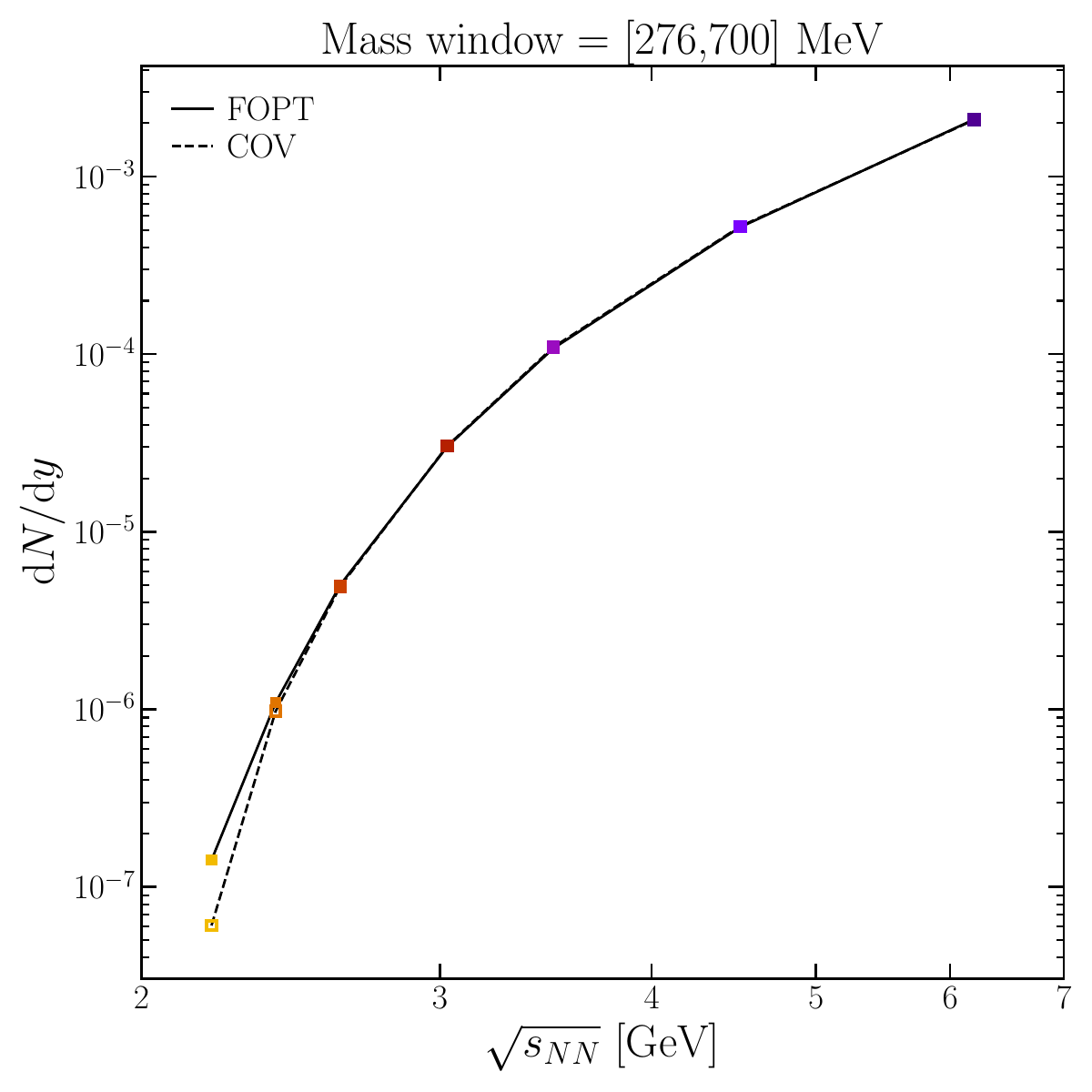}
\includegraphics[width=0.49\textwidth]{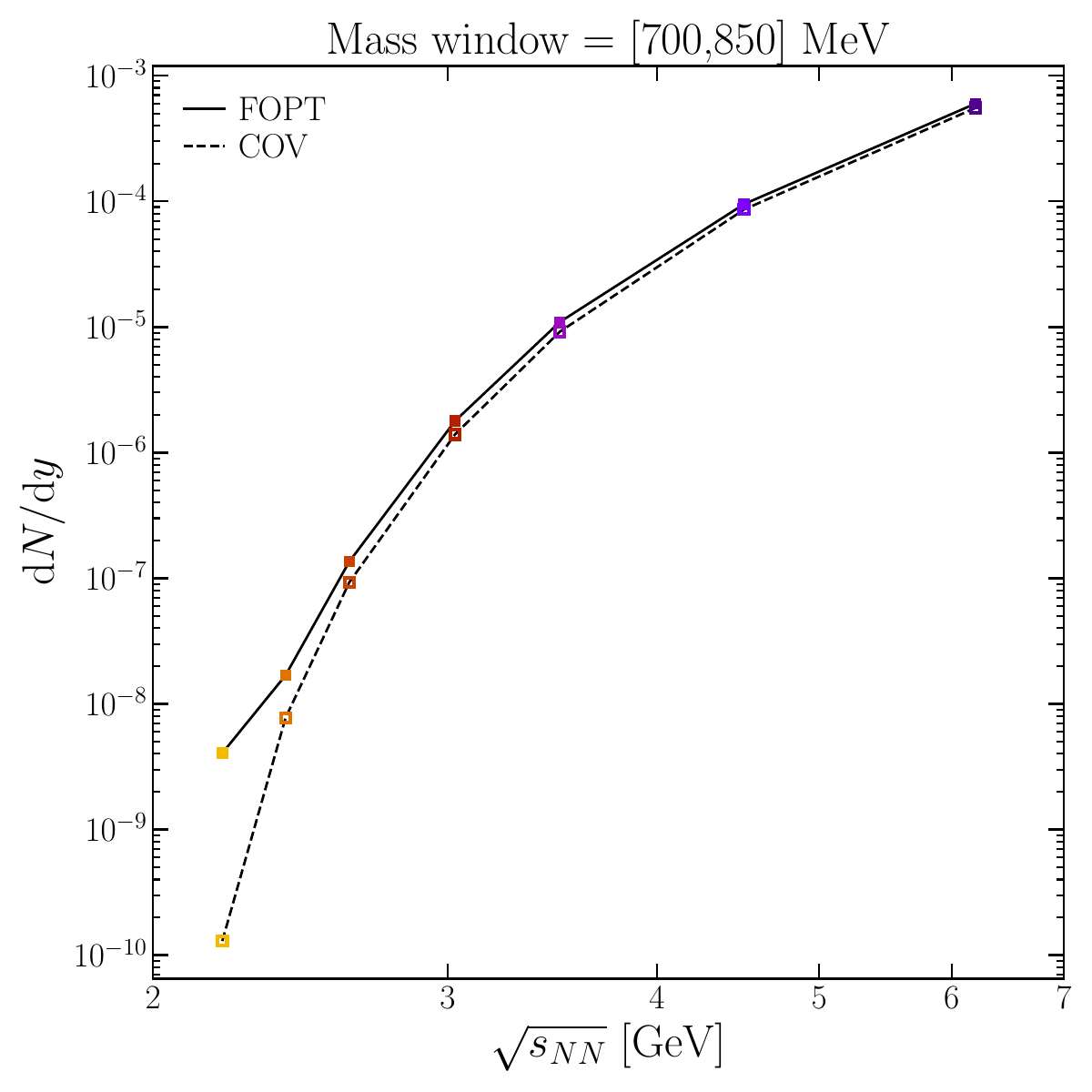}
\caption{Excitation function of $\mathrm{d}N/\mathrm{d}y$ versus $\sqrt{s_{NN}}$ for (top) low-mass $M\in[276,700]~\mathrm{MeV}$ and (bottom) pole-mass $M\in[700,850]~\mathrm{MeV}$.}
\label{fig:dndy}
\end{figure}

Figure~\ref{fig:dndy} shows the midrapidity dilepton yields in the low-mass and pole-mass windows as a function of $\sqrt{s_{\rm NN}}$. Solid symbols and lines correspond to the FOPT scenario, whereas open symbols and dashed lines denote the COV scenario. The excitation function demonstrates that the FOPT remains identifiable even after integration over both the fireball evolution and the respective invariant mass window. Both windows show an increasing dilepton yield with energy owing to the larger temperatures, longer-lived fireballs, and increased emission volumes. Superimposed on this common trend is the characteristic FOPT enhancement, which becomes progressively stronger toward lower center-of-mass energies where the trajectories probe the first-order region.
The low-mass window exhibits a noticeable FOPT enhancement only at $2.20~\mathrm{GeV}$. As shown in the time-resolved analysis, this contribution originates predominantly from the reheating stage and accumulates over an extended period. In contrast, the pole-mass window retains a separation between the FOPT and COV scenarios over the entire energy range in which the trajectories encounter the first-order region ($2.20~\mathrm{GeV}$ to $3.50~\mathrm{GeV}$). Because this enhancement is generated during the full non-equilibrium evolution, it survives the subsequent integration over the fireball history more effectively than the late-time low-mass contribution.

The results demonstrate that the characteristic signatures of a FOPT are not washed out by integrating over the full collision history. Experimentally, such a behavior would appear as a modification of the excitation function itself. In particular, the FOPT scenario predicts a pronounced flattening of the pole-mass dilepton yield toward the lowest collision energies, whereas the crossover excitation function continues to decrease rapidly. A beam-energy scan with sufficiently fine energy spacing could therefore search for an anomalous change of slope or curvature in the pole-mass dilepton excitation function.
This signal provides a practical benchmark for future measurements at HADES, FAIR-CBM, and the RHIC Beam Energy Scan. 

\section{Discussion and Conclusion}
\label{sec:discussion}

In the present work, we have combined N$\chi$FD with microscopic in-medium vector-meson self-energies to isolate the effect of a FOPT in the underlying EoS on dilepton radiation in heavy-ion collisions.
This work demonstrates that a FOPT leaves characteristic signatures in dilepton production that can be distinguished from conventional hadronic in-medium broadening. A FOPT enhances dilepton radiation through the non-equilibrium evolution of the fireball, including supercooling, reheating, and delayed freeze-out. The present work combines these dynamics with microscopic, resonance-driven in-medium $\rho$ and $\omega$ self-energies constructed from forward-scattering amplitudes. This provides a unified description in which both the bulk evolution and the vector-meson spectral functions evolve consistently along the dynamical trajectories.

Time-resolved emission rates reveal that the first-order transition produces two complementary signatures. The pole-mass region is modified during the early non-equilibrium evolution, whereas the low-mass continuum receives an additional contribution during the reheating stage. These distinct emission histories provide the physical origin of the spectral modifications observed in the integrated dilepton spectra.

The excitation functions demonstrate that the FOPT remains identifiable even after integrating over the complete fireball evolution and selected invariant mass windows. Although the low-mass enhancement is substantially reduced by this integration, the pole-mass window retains a pronounced FOPT–COV separation over the beam-energy range where the trajectories probe the first-order region. Among the observables investigated here, the pole-mass excitation function therefore provides the strongest sensitivity to the underlying phase structure. By keeping the microscopic vector-meson self-energies identical and varying only the macroscopic phase structure, the study demonstrates that EoS dynamics leaves experimentally observable dilepton signatures that can be separated from conventional hadronic in-medium broadening.

The present calculation provides a proof of principle based on a simplified longitudinal Bjorken expansion. The quantitative magnitude of the enhancement, its precise onset energy, and the sharpness of the structure in the excitation function should therefore not be regarded as model-independent predictions. A full 3+1D expansion with transverse dynamics and spatially varying thermodynamic conditions can be expected to smear the structures obtained in the present homogeneous description. The physical mechanism responsible for the enhancement, however, is more general than the particular expansion scenario considered here. Supercooling and delayed conversion of a metastable phase increase the time spent in the emitting medium, whereas the subsequent relaxation of the order parameter can release energy and reheat the system. These are generic consequences of sufficiently strong nonequilibrium first-order phase-transition dynamics and are not specific to the Bjorken geometry. We therefore expect an enhancement of electromagnetic radiation associated with delayed phase conversion to persist qualitatively in more realistic dynamical descriptions, although its magnitude and detailed beam-energy dependence may change.

The present results suggest experimental benchmarks for future dilepton measurements. In particular, the beam-energy dependence of the dilepton yield in the pole-mass region emerges as the most sensitive observable considered here for identifying modifications associated with first-order transition dynamics. Measurements by HADES, the future CBM experiment at FAIR, and the RHIC Beam Energy Scan could therefore test whether the predicted enhancement at low beam energies is realized in nature. Since both dilepton production and HBT interferometry respond to the space-time evolution of the fireball, future studies could investigate correlations between the dilepton enhancement predicted here and changes in HBT radii~\cite{Lisa:2005dd,Li:2008qm}. Such a combined analysis may provide additional constraints on the dynamics of a FOPT.

\begin{acknowledgements}
This work was supported by Suranaree University of Technology (SUT) and has received funding support from the NSRF via the Research and Innovation Acceleration Agency for Competitiveness and Area Development (RCAD) (Program Management Unit for Frontier Brainpower and Future Industries) [grant number B39G690072]. The authors thank Marcus Bleicher and Jan Steinheimer for fruitful discussions on spectral functions.
\end{acknowledgements}

\begin{table*}[htbp]
\centering
\begin{tabular}{l|cccc|cccc}
\hline
\multirow{2}{*}{Particle} & \multicolumn{4}{c|}{UrQMD} & \multicolumn{4}{c}{Eletsky et al.} \\
& $\rho\pi$ & $\rho N$ & $\omega\pi$ & $\omega N$
& $\rho\pi$ & $\rho N$ & $\omega\pi$ & $\omega N$ \\
\hline
N(1520)        &        &        &        &        &        & 0.13 &        &        \\
N(1535)        &        & 0.10 &        &        &        &        &        &        \\
N(1650)        &        & 0.17 &        &        &        &        &        &        \\
N(1675)        &        & 0.01 &        &        &        &        &        &        \\
N(1680)        &        & 0.10 &        &        &        &        &        &        \\
N(1700)        &        & 0.38 &        & 0.22 &        & 0.13 &        &        \\
N(1710)        &        & 0.17 &        & 0.03 &        &        &        &        \\
N(1720)        &        & 0.01 &        & 0.26 &        & 0.87 &        &        \\
N(1875)        &        & 0.46 &        & 0.20 &        &        &        &        \\
N(1880)        &        & 0.32 &        & 0.20 &        &        &        &        \\
N(1895)        &        & 0.32 &        & 0.28 &        &        &        &        \\
N(1900)        &        & 0.33 &        & 0.10 &        & 0.44 &        & 0.30 \\
N(2000)        &        &        &        &        &        & 0.60 &        &        \\
N(2060)        &        & 0.19 &        & 0.04 &        &        &        &        \\
N(2080)        &        &        &        &        &        & 0.46 &        & 0.20 \\
N(2090)        &        &        &        &        &        & 0.32 &        & 0.28 \\
N(2100)        &        & 0.53 &        & 0.17 &        & 0.27 &        &        \\
N(2120)        &        & 0.01 &        & 0.12 &        &        &        &        \\
N(2190)        &        & 0.06 &        & 0.14 &        & 0.11 &        & 0.20 \\
$\Delta$(1620) &        & 0.27 &        &        &        &        &        &        \\
$\Delta$(1700) &        & 0.27 &        &        &        & 0.08 &        &        \\
$\Delta$(1900) &        & 0.41 &        &        &        & 0.38 &        &        \\
$\Delta$(1905) &        & 0.26 &        &        &        & 0.26 &        &        \\
$\Delta$(1940) &        &        &        &        &        & 0.86 &        &        \\
$\Delta$(2000) &        &        &        &        &        & 0.22 &        &        \\
$\phi$~(1020)   & 0.15 &        &        &        & 0.13 &        &        &        \\
$a_1$(1260)    & 0.68 &        &        &        & 0.68 &        &        &        \\
$a_2$(1320)    & 0.70 &        & 0.11 &        & 0.70 &        & 0.10 &        \\
$h_1$(1170)    & 1.00 &        &        &        & 1.00 &        &        &        \\
$\omega$~(1420) & 0.30 &        & 0.70 &        & 1.00 &        &        &        \\
$\rho$~(1700)   & 1.00 &        &        &        &        &        &        &        \\
$\pi$~(1300)    & 1.00 &        &        &        & 0.32 &        &        &        \\
$\pi_2$(1670)  & 0.31 &        & 1.00 &        &        &        &        &        \\
$b_1$(1235)    &        &        & 1.00 &        &        &        & 1.00 &        \\
$a_0$(1450)    &        &        & 1.00 &        &        &        &        &        \\
$\rho_3$(1690) &        &        & 0.16 &        &        &        &        &        \\
\hline
\end{tabular}
\caption{Merged resonance branching ratios for $\rho$ and $\omega$ channels: UrQMD particle table content vs.\ the resonance set used in the original Eletsky et al. implementation~\cite{Eletsky:2001bb}. Empty entries indicate channels not present in the corresponding set.}
\label{tab:merged_urqmd_eletsky}
\end{table*}


\bibliography{references}
\bibliographystyle{spphys}
\end{document}